\documentclass[a4paper,prd,showkeys,nofootinbib]{revtex4}
\usepackage[T1]{fontenc}
\usepackage[latin1]{inputenc}
\usepackage{graphicx}
\usepackage{amsmath}
\usepackage{amssymb}

\newtheorem{theorem}{Theorem}
\newtheorem{lemma}{Lemma}
\DeclareMathOperator{\diag}{diag}
\DeclareMathOperator{\tr}{tr}

\begin{document}

\title{On the Weyl Tensor Classification in All Dimensions and its Relation with Integrability Properties}

\author{Carlos  Batista}
\email{carlosbatistas@df.ufpe.br}
\affiliation{Departamento de F\'{\i}sica, Universidade Federal de Pernambuco, 50670-901
Recife - PE, Brazil}


\begin{abstract}
In this paper the Weyl tensor is used to define operators that act on the space of forms. These operators are shown to have interesting properties and are used to classify the Weyl tensor, the well known Petrov classification emerging as a special case. Particularly, in the Euclidean signature this classification turns out be really simple. Then it is shown that the integrability condition of maximally isotropic distributions can be described in terms of the invariance of certain subbundles under the action of these operators. Here it is also proved a new generalization of the Goldberg-Sachs theorem, valid in all even dimensions, stating that the existence of an integrable maximally isotropic distribution imposes restrictions on the optical matrix. Also the higher-dimensional versions of the self-dual manifolds are investigated. These topics can shed light on the integrability of Einstein's equation in higher dimensions.
\end{abstract}

\maketitle
\section{Introduction }
The Petrov classification \cite{Petrov,art1} is a scheme to classify the Weyl tensor in four dimensions that has been responsible for much progress on general relativity. In particular it was of fundamental importance for the discovery of one of the most important solutions to Einstein's equation, the Kerr metric \cite{Kerr}. The main reason behind the usefulness of such classification is encoded on the Goldberg-Sachs(GS) theorem, which states that in a vacuum(Ricci-flat) four-dimensional Lorentzian manifold the Weyl tensor is algebraically special if, and only if, the spacetime admits a congruence of null geodesics that is shear-free \cite{Goldberg-Sachs}. Later it has been proved that this theorem could be extended to all signatures if the concept of shear-free geodesics is replaced by the existence of an integrable maximally isotropic distribution \cite{Plebanski2,art2}, revealing the geometrical content of the GS theorem.

The intent of the present article is to generalize the Petrov classification to all dimensions and to find extensions of Goldberg-Sachs theorem (specially in even dimensions). Hopefully this will help the search of new exact solutions to Einstein's equation in higher dimensions, with relevance to string theory compactifications. Since Einstein's equation is non-linear there is usually no hope to find all its solutions, however with the help of GS theorem Kinnersley was able to find all Petrov type D vacuum solutions in four dimensions \cite{typeD - Kinnersley}, a really impressive achievement. The topics discussed here can, analogously, help to fully integrate Einstein's equation under certain circumstances in higher dimensions. Also the results obtained here should promote more geometrical insight about the Weyl tensor with possible applications in general relativity. Moreover this work has applicability in differential geometry, specially in areas related to the integrability of distributions and in spinor geometry, since maximally isotropic distributions are ubiquitous in this article. Finally it is pertinent to mention that recently it was made an explicit connection between Navier-Stokes' and Einstein's equations \cite{Navier-Stokes}, in this connection the classification of the Weyl tensor makes a prominent role.

Different classification schemes for the Weyl tensor in dimensions greater than four have been already defined during the last decade. In \cite{CMPP} it was put forward a form to classify any tensor in Lorentzian spaces based on the so called boost weight, the well known CMPP classification. Extensions of the GS theorem using such classification were looked for and some progress has been made \cite{GS-CMPP,GS-Ortaggio12,M. Ortaggio-Robinson-Trautman}. A recent review of this approach is available in reference \cite{CMPP-Review}. In \cite{HigherGSisotropic1,HigherGSisotropic2} it was defined a classification scheme for the Weyl tensor valid in any dimension and signature based on the maximally isotropic structures and it was found a generalization of the GS theorem that will be used in the present article. In \cite{5D classification} the Weyl tensor was classified in five dimensions using spinor techniques and applications were made. In \cite{Spin6D} it was worked out a classification scheme for the Weyl tensor in six dimensions using spinors and the generalized GS theorem of \cite{HigherGSisotropic2} was elegantly translated to the spinor language. Finally, a very recent paper used spinorial language to define a classification for the Weyl tensor in even dimensions \cite{Taghavi-Spinors}. The classification presented here is more refined than the previous ones and has the classification of reference \cite{Spin6D} and the Petrov classification as special cases. A short and nice review of the previous literature can be found in \cite{Reall - Review}, where some possible applications of these subjects are also discussed.

In section \ref{Sec- C_p} the Weyl tensor is used to define operators $\textbf{C}_p$ that act on the space of $p$-forms, it is also shown that such operators have nice properties. For example, in the Euclidean signature they are Hermitean. These operators are then used to classify the Weyl tensor in any dimension. Section \ref{Sec- EvenDim} shows that in even dimensions, $d=2n$, the operator $\textbf{C}_n$ preserves the space of (anti-)self-dual $n$-forms and this is used to split this operator into the direct sum of a self-dual part and an anti-self-dual part.  In section \ref{Sec- MaximallyIstrop} it is proved that in even dimensions the integrability condition for maximally isotropic distributions found in \cite{HigherGSisotropic2} can be nicely expressed in terms of the operator $\textbf{C}_n$. Section \ref{Sec- SimpleNot} introduces a notation in which the map $\textbf{C}_p$ takes a really simple and elegant form. Although not deeply exploited here this notation seems to be promising. Then section \ref{Sec- OtherClassif.} gives a brief review of some other important methods of classification for the curvature and argues how the classification introduced here can be helpful.  Section \ref{Sec- Mariot-Rbinson} provides a new and simple proof of the generalized Mariot-Robinson theorem that will be important for future sections. Finally, in sections \ref{Sec- OpticalMatrix} and \ref{Sec- GSgeneralized} are shown links between the integrability of null structures and the optical matrix. In particular it is proved a new generalization of the GS theorem, stating that in even dimensions the existence of integrable maximally isotropic distributions imposes restrictions on the optical matrix. Appendix \ref{Appendix-Segre} defines a refined version of the Segre classification that is used to classify the operators $\textbf{C}_p$, while appendix \ref{Appendix-SimpleForms} presents important results about simple forms that are used throughout the article.

In the present work the vector bundles are assumed to be complexified so that the results can be applied to any signature, on the same lines of references \cite{art1,art2,Spin6D}. All results obtained here are local and it is always assumed that the manifold is endowed with a non-degenerate metric and its Levi-Civita connection. Throughout the article some examples are worked out with the intent of facilitating the comprehension and showing possible applications of the concepts and tools introduced here.

\section{The Map $\textbf{C}_p$}\label{Sec- C_p}
Let $(M,g)$ be a manifold of dimension $d$ endowed with a non-degenerate real metric tensor $g_{\mu\nu}$ of arbitrary signature $s\,$\footnote{In this article the metric tensor is assumed to be real just because sometimes it is convenient to deal with a real Weyl tensor. But almost all results presented here are also valid for the case of a complex metric.}. When convenient the tangent bundle of this manifold will be implicitly assumed to be complexified. Since all results throughout this article will be local we can always assume the existence of a volume form denoted by $\varepsilon_{\nu_1\nu_2\ldots\nu_d}$. This $d$-form obeys to the following well known identity\footnote{As usual the indices enclosed by square brackets are anti-symmetrized, while the indices inside round brackets are symmetrized. For example, $T_{[ab]}=\frac{1}{2}(T_{ab}-T_{ba})$ and $T_{(ab)}=\frac{1}{2}(T_{ab}+T_{ba})$. Also, repeated indices are assumed to be summed.}:
\begin{equation}\label{EpsilonEps}
  \varepsilon^{\mu_1\ldots\mu_p\,\nu_{p+1}\ldots\nu_d}\,\varepsilon_{\mu_1\ldots\mu_p\,\alpha_{p+1}\ldots\alpha_d}\;=\;p!(d-p)!\, (-1)^{\frac{d-s}{2}} \delta_{\alpha_{p+1}}^{\;[\nu_{p+1}}\ldots\delta_{\alpha_{d}}^{\;\nu_{d}]}\,.
\end{equation}
Given a $p$-form $K_{\nu_1\ldots\nu_p}=K_{[\nu_1\ldots\nu_p]}$ then its Hodge dual is the $(d-p)$-form defined by:
\begin{equation}\label{Hdge Dual}
 \widetilde{K}_{\mu_1\ldots\mu_{d-p}}\;=\; \frac{1}{p!} \,\varepsilon^{\nu_1\ldots\nu_{p}}_{\phantom{\nu_1\ldots\nu_{p}}\mu_1\ldots\mu_{d-p}} \, K_{\nu_1\ldots\nu_p}\,.
\end{equation}
Taking the Hodge dual twice and using equation (\ref{EpsilonEps}) it is straightforward to prove the following important relation:
\begin{equation}\label{Double Dual}
 \widetilde{\widetilde{K}}_{\nu_1\ldots\nu_{p}} \;=\; (-1)^{[p(d-p)+ \frac{d-s}{2}]}\, K_{\nu_1\ldots\nu_{p}}\,.
\end{equation}

The tangent bundle of $(M,g)$ is assumed to be endowed with the Levi-Civita connection, whose trace-less part of the curvature, called the Weyl tensor, will be denoted by $C_{\mu\nu\alpha\beta}$. This tensor has the following symmetries:
\begin{equation}\label{Weyl Symmetries}
  C_{\mu\nu\alpha\beta}=C_{[\mu\nu][\alpha\beta]}=C_{\alpha\beta\mu\nu}\;\;;\;\; C_{\mu[\nu\alpha\beta]}=0\;\;;\;\; C^\mu_{\phantom{\mu}\nu\mu\beta}=0\,.
\end{equation}
By means of the Weyl tensor we can define the following linear operator that act on the space of $p$-forms with $p\geq2$:
\begin{equation}\label{C_p}
  \textbf{C}_p:\quad\; K_{\nu_1\ldots\nu_p}\; \mapsto \; K'_{\nu_1\ldots\nu_p} = C^{\alpha\beta}_{\phantom{\alpha\beta}[\nu_1\nu_2}K_{\nu_3\ldots\nu_{p}] \alpha\beta}\,.
\end{equation}

The particular case $p=2$ of the above operator has been of fundamental importance to the development of general relativity in four dimensions, since it gives rise to the well known Petrov classification \cite{art1,Petrov}. Such classification has been used to find very important solutions to Einstein's equation, the most important examples being the Kerr metric \cite{Kerr} and all type $D$ vacuum solutions \cite{typeD - Kinnersley}. The operator $\textbf{C}_2$ was also investigated in higher dimensions on reference \cite{BivectHighDim}, with the intent of refining CMPP classification. It is also interesting to note that in six dimensions the operator $\textbf{C}_3$ reduces to the Weyl operator defined on \cite{Spin6D} using spinors. But, as far as the present author knows, the operators $\textbf{C}_p$, for arbitrary $p$, have not been defined before. The goal of the present article is to study some properties of the maps $\textbf{C}_p$ for $p>2$, define a classification for the Weyl tensor based on them and, in some cases, relate these maps to a generalization of the Goldberg-Sachs theorem.

 Now let us work out a useful relation between the Weyl operator $\textbf{C}_p$ and the Hodge dual map. If $K'$ is the image of the $p$-form $K$ under the operator $\textbf{C}_p$, as in equation (\ref{C_p}), then we have:
 \begin{gather*}\label{Dual of C_p}
   \widetilde{K'}^{\nu_1\ldots\nu_{(d-p)}}\; =\; \frac{1}{p!}\, \varepsilon^{\mu_1\ldots\mu_{p} \, \nu_1\ldots\nu_{d-p}} C^{\alpha\beta}_{\phantom{\alpha\beta}\mu_1\mu_2}\, \widetilde{\widetilde{K}}_{\mu_3\ldots\mu_{p} \alpha\beta}\,(-1)^{[(d-p)p+ \frac{d-s}{2}]} \;=\; \nonumber\\
   \stackrel{(\ref{Hdge Dual})}{=}\;  \frac{(-1)^{[(d-p)p+ \frac{d-s}{2}]}}{p!\,(d-p)!}\, \varepsilon^{\mu_1\ldots\mu_{p} \, \nu_1\ldots\nu_{d-p}} C^{\alpha\beta}_{\phantom{\alpha\beta}\mu_1\mu_2}\,  \,\varepsilon_{ \sigma_1 \ldots \sigma_{d-p} \mu_3\ldots\mu_{p} \alpha\beta} \,\widetilde{K}^{\sigma_1 \ldots \sigma_{d-p}} \;= \nonumber\\
   \stackrel{(\ref{EpsilonEps})}{=}\; \frac{(p-2)!\,(d-p+2)!}{p!\,(d-p)!} \, \delta_{\alpha}^{\;[\mu_1}\delta_\beta^{\;\mu_2}\delta_{\sigma_1}^{\;\nu_1}\ldots\delta_{\sigma_{d-p}}^{\;\nu_{d-p}]} \, C^{\alpha\beta}_{\phantom{\alpha\beta}\mu_1\mu_2} \widetilde{K}^{\sigma_1 \ldots \sigma_{d-p}} \;\stackrel{(\ref{Weyl Symmetries})}{=}\;  C^{\phantom{\mu_1\mu_2}[\nu_1\nu_2}_{\mu_1\mu_2} \,\widetilde{K}^{\nu_3 \ldots \nu_{d-p}]\mu_1\mu_2}\,.
 \end{gather*}
 Where above it was used equations (\ref{Hdge Dual}), (\ref{C_p}) and (\ref{Double Dual}) in the first equality and the numbers above the other equal signs refer to the used equations. This identity has been already known for the specific case $d=4$ and $p=2$, in which case it has made a prominent role on the development of Petrov classification \cite{Bel}. The above equation is then the generalization of this known fact to all dimensions and to all values of $p$. Such result can be put in a more elegant form if an abstract notation is used.  Denoting the bundle of $p$-forms on $M$ by $\wedge^pM$ then $\textbf{C}_p:\wedge^pM \rightarrow \wedge^pM$ is the abstract operator that implements the map defined in equation (\ref{C_p}), while $\textbf{E}_p:\wedge^pM \rightarrow \wedge^{d-p}M$ will denote the operator that when acts on a $p$-form gives its Hodge dual. In this notation the above result is written in the following form:
\begin{equation}\label{Commutation}
  \textbf{E}_p \, \textbf{C}_p \;=\; \textbf{C}_{d-p}\, \textbf{E}_{p}\,.
\end{equation}

Now we are able to define an algebraic classification scheme for the Weyl tensor valid in all dimensions, we just have to find matrix representations for the operators $\textbf{C}_p$ and use the refined Segre classification (see appendix \ref{Appendix-Segre}). Thus the classification of the Weyl tensor proposed here amounts to gathering the refined Segre types of the operators $\textbf{C}_p$ for all possible values of $p$.  But some of these calculations will be redundant, since by equation (\ref{Double Dual}) we have that the operator $\textbf{E}_p$ is invertible, with inverse proportional to $\textbf{E}_{d-p}$. So using this information in equation (\ref{Commutation}) we have that $\textbf{C}_p=\textbf{E}_p^{\;-1}\textbf{C}_{d-p}\textbf{E}_p$, \textit{i.e.}, the operator $\textbf{C}_p$ can be obtained from $\textbf{C}_{d-p}$ by a similarity transformation, therefore the algebraic type of these operators is the same. Special attention shall be deserved for the cases $p=d$ and $p=(d-1)$ since the ``dual operators'', $\textbf{C}_0$ and $\textbf{C}_1$, are not defined. But it is not difficult to prove that the operators $\textbf{C}_d$ and $\textbf{C}_{d-1}$ are both zero, because of the traceless property of the Weyl tensor. \emph{So it is just necessary to classify the Weyl operators $\textbf{C}_p$ for integer values of $p$ pertaining to the interval $2 \leq p\leq \frac{d}{2}$.}
\newline

If $\textbf{K}_p$ and $\textbf{L}_p$ are $p$-forms then the following symmetric inner product can be defined on the space $\wedge^pM$:
\begin{equation}\label{Inner product}
    <\textbf{K}_p,\textbf{L}_p>\;=\; K^{\mu_1\ldots\mu_p}\,L_{\mu_1\ldots\mu_p}
\end{equation}
Such inner product is non-degenerate and its signature depends on the signature of the metric $g$. If $\{\textbf{B}^i\}$ is a basis for the space $\wedge^pM$ such that $<\textbf{B}^i,\textbf{B}^j>\,=h^{ij}$ then denoting by $h_{ij}$ the inverse matrix of $h^{ij}$ it follows that the $p$-forms $\textbf{B}_i=h_{ij}\textbf{B}^j$ obey to the identity $<\textbf{B}_i,\textbf{B}^j>\,=\delta_i^{\phantom{i}j}$. Thus if $\textbf{K}_p$ is a $p$-form then $\textbf{K}_p=\,<\textbf{B}_i,\textbf{K}_p>\textbf{B}^i$, from which it follows that $\textbf{B}_i^{\;\mu_1\ldots\mu_p} \textbf{B}^i_{\;\nu_1\ldots\nu_p} = \delta_{\nu_{1}}^{\;[\mu_{1}}\ldots\delta_{\nu_{p}}^{\;\mu_{p}]}$. The matrix representation of the operator $\textbf{C}_p$ on the basis $\{\textbf{B}^i\}$ is given by $C_{p\,ij}= \,<\textbf{B}_i,\textbf{C}_p(\textbf{B}^j)>$, so that using the previous results we see that the trace of this operator is zero:
 \begin{equation}\label{TraceZero C_p}
 \textrm{tr}(\textbf{C}_p)=C_{p\,ii}= \textbf{B}_i^{\;\mu_1\ldots\mu_p} C^{\alpha\beta}_{\phantom{\alpha\beta}\mu_1\mu_2}\textbf{B}^i_{\;\mu_3\ldots\mu_p\alpha\beta}= C^{\alpha\beta}_{\phantom{\alpha\beta}\mu_1\mu_2} \delta_{\alpha}^{\;[\mu_{1}}\delta_{\beta}^{\;\mu_{2}}\delta_{\mu_{3}}^{\;\mu_{3}}\ldots\delta_{\mu_{p}}^{\;\mu_{p}]} \propto C^{\alpha\beta}_{\phantom{\alpha\beta}\alpha\beta}=0\,.
 \end{equation}
Also, using the symmetry $C_{\mu\nu\alpha\beta}=C_{\alpha\beta\mu\nu}$ of the Weyl tensor it is simple matter to show that the Weyl operator $\textbf{C}_p$ is self-adjoint with respect to this inner product, \textit{i.e}, the following equation is valid:
\begin{equation}\label{C_p Autoadjunto}
 <\textbf{K}_p,\textbf{C}_p(\textbf{L}_p)> \;=\; <\textbf{C}_p(\textbf{K}_p),\textbf{L}_p>\,.
\end{equation}

In the special case of $(M,g)$ being a real manifold of Euclidean signature it follows that the inner product of $p$-forms defined on equation (\ref{Inner product}) is positive definite. In this case the self-adjointness of operators $\textbf{C}_p$ guarantees that they can be diagonalized and that the eigenvalues are real. This property imposes huge restrictions on the possible algebraic types that the Weyl tensor can have according the classification defined above, since the refined Segre types of operators $\textbf{C}_p$ will depend just on the degeneracy of each eigenvalue and on the dimension of the kernel of these operators. So we have the following lemma:
\begin{lemma}\label{Lemma-EuclideanDiag}
If the metric $g$ has Euclidean signature then the operators $\textbf{C}_p$ admit a traceless diagonal matrix representation with real eigenvalues. In particular this implies that on the refined Segre classification of these operators all numbers inside the square bracket will be 1 (see appendix \ref{Appendix-Segre}).
\end{lemma}

\section{Even Dimensions and the Self-duality}\label{Sec- EvenDim}
In this section the dimension of the manifold $M$ is assumed to be even, $d=2n$. In such case plugging $p=n$ on equation (\ref{Double Dual}) yields $\textbf{E}_n\textbf{E}_n=(-1)^{n^2+n+\frac{s}{2}} \textbf{1}_n=(-1)^{\frac{s}{2}} \textbf{1}_n$, where $\textbf{1}_n$ is the identity operator of space $\wedge^nM$. So the space of $n$-forms can be split into a direct sum of the eigen-spaces of $\textbf{E}_n$, $\wedge^nM = S^+ \oplus S^-$, where $S^+$ is the space of self-dual $n$-forms and $S^-$ is the space of anti-self-dual $n$-forms, defined by:
$$ S^{\pm} = \{\textbf{K}_n\;\in\;\wedge^{n}M \;|\; \textbf{E}_n(\textbf{K}_n)= \pm \epsilon \textbf{K}_n  \}\,.    $$
Where $\epsilon$ is equal to $1$ or $i$ depending on whether $\frac{s}{2}$ is respectively even or odd. The spaces $S^+$ and $S^-$ both have the same dimension, $\frac{1}{2}\left(^{2n}_{\,\,n}\right)$. Using equations (\ref{Hdge Dual}) and (\ref{Inner product}) a simple index rearrangement shows that the following property holds:
\begin{equation*}
 <\textbf{K}_n\,,\,\textbf{E}_n(\textbf{L}_n)> \;= \; (-1)^{n^2}\, <\textbf{E}_n(\textbf{K}_n)\,,\,\textbf{L}_n>\,.
\end{equation*}
Where $\textbf{K}_n$ and $\textbf{L}_n$ are arbitrary $n$-forms. Using this equation we arrive at the following result:
\begin{lemma}\label{Lemma-S^+- orthogonal}
In a manifold of dimension $d=2n$ if $n$ is even then the spaces $S^+$ and $S^-$ are orthogonal to each other with respect to the inner product $<,>$. On the other hand if $n$ is odd then every element of $S^+$ is orthogonal to every element of $S^+$ and every element of $S^-$ is orthogonal to every element of $S^-$.
\end{lemma}


Now let us see that the operator $\textbf{C}_n$ has the important property of preserving the spaces $S^\pm$. If $\textbf{K}_n^\pm$ pertain to $S^\pm$ and $\textbf{L}_n^\pm=\textbf{C}_n(\textbf{K}_n^\pm)$ then equation (\ref{Commutation}) implies that:
$$ \textbf{E}_n(\textbf{L}_n^\pm ) = \textbf{E}_n \textbf{C}_n(\textbf{K}_n^\pm) = \textbf{C}_n\textbf{E}_n(\textbf{K}_n^\pm) = \textbf{C}_n(\pm \epsilon \textbf{K}_n^\pm) = \pm \epsilon  \textbf{L}_n^\pm \,.$$
Therefore it is useful to define the following operators:
\begin{equation}\label{C^+-}
    \textbf{C}^\pm:\, \wedge^nM\rightarrow \wedge^nM \;\;,\;\;\;       \textbf{C}^\pm \; \equiv \; \frac{1}{2}(\textbf{C}_n \,\pm \,\epsilon^{-1}\,\textbf{C}_n\textbf{E}_n )\,.
\end{equation}
It is immediate to see that $\textbf{C}^\pm$ is zero when restricted to $S^\mp$, \textit{i.e.}, $\textbf{C}_n = \textbf{C}^+  \oplus \textbf{C}^-$. Thus the matrix representation of $\textbf{C}_n$ can be put in block-diagonal form with two blocks of the same dimension. This property restrict enormously the possible algebraic types that the operator $\textbf{C}_n$ can have. This has been already known in four dimensions and was of fundamental importance to the development of the Petrov classification \cite{Petrov,art1}. Also this has been noted in six dimensions using spinor calculus \cite{Spin6D}. This property of $\textbf{C}_n$ guarantees that when convenient the operators $\textbf{C}^\pm$ can be assumed to act on $S^\pm$ instead of $\wedge^nM$.

Let us note that the trace of both operators $\textbf{C}^+$ and $\textbf{C}^-$ is zero. From equation (\ref{TraceZero C_p}) it follows that the trace of $\textbf{C}_n$ is zero, so that using the definition of operators $\textbf{C}^\pm$, equation (\ref{C^+-}), we see that $\tr(\textbf{C}^\pm)=\pm\epsilon^{-1}\tr(\textbf{C}_n\textbf{E}_n )$. Using steps similar to the ones used to evaluate $\tr(\textbf{C}_p)$ we find for $n\geq2$:
$$ \tr(\textbf{C}_n\textbf{E}_n )= \frac{1}{n!}\textbf{B}_i^{\;\mu_1\ldots\mu_n} C^{\alpha\beta}_{\phantom{\alpha\beta}\mu_1\mu_2} \varepsilon^{\nu_1\ldots\nu_n}_{\phantom{\nu_1\ldots\nu_n}\mu_3\ldots\mu_n\alpha\beta} \textbf{B}^i_{\;\nu_1\ldots\nu_n}=
\frac{1}{n!}C^{\alpha\beta}_{\phantom{\alpha\beta}\mu_1\mu_2} \varepsilon^{\nu_1\ldots\nu_n}_{\phantom{\nu_1\ldots\nu_n}\mu_3\ldots\mu_n\alpha\beta}
\delta_{[\nu_1}^{\;\mu_{1}}\ldots\delta_{\nu_{n}]}^{\;\mu_{n}}=0\,.$$
%
Where the last equality follows from the Bianchi identity, $C_{[\alpha\beta\mu]\nu}=0$. So the operators $\textbf{C}^\pm$ have vanishing trace.

From lemma \ref{Lemma-S^+- orthogonal} and from the fact that the inner product $<,>$ in $\wedge^nM$ is non-degenerate we have that when $n$ is odd it is possible to introduce a basis $\{\textbf{B}^{+i}\}$ for $S^+$ and a basis $\{\textbf{B}^{-i}\}$ for $S^-$ such that $<\textbf{B}^{+i},\textbf{B}^{-j}>=\delta^{ij}$. To see this just start with a basis for $S^+$ and a basis for $S^-$ and then use the Gram-Schmidt process to conveniently redefine the basis of $S^-$. The matrix representation of $\textbf{C}^+$ on the basis $\{\textbf{B}^{+i}\}$ is then $C^+_{ij}= <\textbf{B}^{-i},\textbf{C}_n(\textbf{B}^{+j})>$, while the matrix representation of $\textbf{C}^-$ on the basis $\{\textbf{B}^{-i}\}$  is $C^-_{ij}= <\textbf{B}^{+i},\textbf{C}_n(\textbf{B}^{-j})>$. Then from equation (\ref{C_p Autoadjunto}) it follows that $C^+_{ij}=C^-_{ji}$. So when $n$ is odd it follows that operator $\textbf{C}^+$ can be obtained from $\textbf{C}^-$ and \textit{vice versa}. On the other hand when $n$ is even the operators $\textbf{C}^+$ and $\textbf{C}^-$ are in principle independent of each other. The results obtained so far can be summarized by the following theorem.

\begin{theorem}\label{Theor. C^+andC^-}
When the dimension of the manifold is $d=2n$ the operator $\textbf{C}_n$ is the direct sum of an operator that acts on the space of self-dual $n$-forms,  $\textbf{C}^+: S^+\rightarrow S^+$, and an operator that acts  on the space of anti-self-dual $n$-forms, $\textbf{C}^-: S^-\rightarrow S^-$. Both operators, $\textbf{C}^+$ and $\textbf{C}^-$, have vanishing trace. In the particular case of odd $n$ it follows that the operator $\textbf{C}^-$ is the adjoint of the operator $\textbf{C}^+$, \textit{i.e.}, in a suitable basis the matrix representation of $\textbf{C}^-$ is the transpose of the matrix that represents $\textbf{C}^+$.  \end{theorem}

This theorem implies that when $n$ is odd then to classify $\textbf{C}_n$ is equivalent to classify $\textbf{C}^+$, since automatically the operator $\textbf{C}^-$ has the same algebraic type of $\textbf{C}^+$. In the case $n=3$ this was already proved using spinorial techniques on reference \cite{Spin6D}. In turn, when $n$ is even classify $\textbf{C}_n$ is equivalent to compute the algebraic types of both operators $\textbf{C}^+$ and $\textbf{C}^-$.

In four dimensions a manifold is called self-dual when $\textbf{C}^-=0$ and $\textbf{C}^+\neq0$. These manifolds have been widely studied in the past \cite{Plebanski3}, in particular it has been shown that Einstein's vacuum equation on self-dual manifolds reduces to a single second-order differential equation \cite{Plebanski3}. Now it is natural to ask wether the notion of self-dual manifolds can be extended to higher dimensions. Theorem \ref{Theor. C^+andC^-} says that if the dimension is $d=2n$ with odd $n$ then $\textbf{C}^-=0$ implies $\textbf{C}^+=0$, so that the Weyl tensor is identically zero. But in principle if the dimension is multiple of four, even $n$, the operators $\textbf{C}^+$ and $\textbf{C}^-$ are independent of each other so that the self-dual manifolds can be defined. But it turns out that a laborious investigation of the self-dual constraint in 8 dimensions reveals that if $\textbf{C}^+=0$ then $\textbf{C}^-=0$. Thus arriving at the following lemma:

\begin{lemma}\label{Lemma-Self-dual-man.}
If the dimension of the manifold is $d=2n$ with odd $n$ or $n=4$ then the constraint $\textbf{C}^+=0$ implies that the whole Weyl tensor vanishes. So in these dimensions the notion of self-dual manifold cannot be introduced.
\end{lemma}
Although in the case of even $n$ this result has been worked out by the author only for the case $n=4$ the calculations seems to indicate that lemma \ref{Lemma-Self-dual-man.} is valid for all $n\geq4$. Thus, concerning the Weyl tensor, the dimension four appears to be very special, since in this dimension the operators $\textbf{C}^+$ and $\textbf{C}^-$ can be independent of each other \cite{art1}. Now in order to get acquainted with the tools introduced so far let us see how the Petrov classification emerges in the present formalism.

\begin{quote}
 \textbf{Example: } In four dimensions, $d=4$, the operators $\textbf{C}_3$ and $\textbf{C}_4$ are trivially zero, as explained in section \ref{Sec- C_p}. So we can only use $\textbf{C}_2$ to algebraically classify the Weyl tensor. According to theorem \ref{Theor. C^+andC^-} it follows that $\textbf{C}_2 =\textbf{C}^+ \oplus \textbf{C}^-$, where $\textbf{C}^+$ is independent from $\textbf{C}^-$, and both have vanishing trace. Since the space of 2-forms has six dimensions it follows that the operators $\textbf{C}^\pm$ act on 3-dimensional spaces, $S^\pm$. Then using the refined Segre classification (appendix \ref{Appendix-Segre}) it follows that the possible types for $\textbf{C}^\pm$ are $[1,1,1|]$, $[2,1|]$, $[(1,1),1|]$, $[|3]$, $[|2,1]$  and $[|1,1,1]$, these types are respectively called $I$, $II$, $D$, $III$, $N$ and $O$ (see reference \cite{art1}). Note that the refined Segre type $[1,1|1]$ is also possible and in the Petrov classification it also corresponds to the type $I$, so that the refined Segre classification provides one more algebraic type than the Petrov classification.

\quad Thus the operator $\textbf{C}^+$ can have six Petrov types, just as $\textbf{C}^-$. Then the Weyl tensor can have $6\times6=36$ Petrov types. But from an intrinsic point of view just $21$ types are different, because the spaces $S^+$ and $S^-$ are interchanged by a simple multiplication of the volume form, $\boldsymbol{\varepsilon}$,  by $-1$. When the Weyl tensor is real and the signature Euclidean the operators $\textbf{C}^\pm$ admit diagonal representations, thus $\textbf{C}^\pm$ can just have types $I$, $D$ or $O$. While if the Weyl tensor is real and the signature Lorentzian then the operators $\textbf{C}^\pm$ can have any type, but the type of $\textbf{C}^+$ must be the same of $\textbf{C}^-$ \cite{art1}.
\end{quote}

Before proceed some comments about reality conditions are in order. For a manifold $(M,g)$ of dimension $d=2n$ and signature $s$ if $(s/2)$ is even then the eigenvalues of $\textbf{E}_n$ are real, $\pm1$. In this case it is possible to find real bases for both spaces $S^+$ and $S^-$. On the other hand if $(s/2)$ is odd the eigenvalues of $\textbf{E}_n$ are $\pm i$ so that the elements of $S^+$ and $S^-$ must be complex. In this last case if an $n$-form is self-dual then its complex-conjugate will be anti-self-dual, and \textit{vice versa}. It is also important to note that if the Weyl tensor is real, as assumed in the present article, then this reality condition generally constrains the possible algebraic types of $\textbf{C}_n$. These constraints depend on the signature of the metric, just as happens to $\textbf{C}_2$ in four dimensions \cite{art1}.

Particularly in the Euclidean signature we have $s=d=2n$, so that if $n$ is even then it is possible to define real bases $\{\textbf{B}^{+i}\}$ and $\{\textbf{B}^{-i}\}$ for $S^+$ and $S^-$ respectively such that $<\textbf{B}^{+i},\textbf{B}^{+j}>=\delta^{ij}= <\textbf{B}^{-i},\textbf{B}^{-j}>$ (see lemma \ref{Lemma-S^+- orthogonal}). Using equation (\ref{C_p Autoadjunto}) we see that in these bases the operators $\textbf{C}^\pm$ have matrix representations that are real and symmetric. In turn, if the signature is Euclidean and $n$ is odd it is always possible to find a basis $\{\textbf{B}^{+i}\}$ for $S^+$ such that $\{\overline{\textbf{B}^{+i}}\}$ is a basis for $S^-$ and  $<\textbf{B}^{+i},\overline{\textbf{B}^{+j}}>=\delta^{ij}$. The basis $\{\textbf{B}^{+i}\}$ can also always be arranged to be such that the matrix representation of $\textbf{C}^+$ is diagonal with real eigenvalues. To prove this note that if $\{\textbf{Q}^{+i}\}$ is a basis for $S^+$ then, since $(s/2)$ is assumed to be odd, the complex conjugate of this basis will be a basis for $S^-$. Now the positive definiteness of $g$ implies that $<\textbf{Q}^{+i},\overline{\textbf{Q}^{+i}}>$ is greater than zero so that by the Gram-Schmidt process we can find a basis $\{\textbf{B}'^{+i}\}$ such that $<\textbf{B}'^{+i},\overline{\textbf{B}'^{+j}}>=\delta^{ij}$. In this basis, because of equation (\ref{C_p Autoadjunto}), the matrix representation of $\textbf{C}^+$ is easily seen to be Hermitean, thus by an unitary transformation we can change the basis $\{\textbf{B}'^{+i}\}$ to another basis $\{\textbf{B}^{+i}\}$ such that $<\textbf{B}^{+i},\overline{\textbf{B}^{+j}}>=\delta^{ij}$ and in which the representation of $\textbf{C}^+$ takes a diagonal form.

\section{Integrability of Maximally Isotropic Distributions and the Operator $\textbf{C}_n$}\label{Sec- MaximallyIstrop}
The celebrated Goldberg-Sachs theorem in its generalized version states that in a Ricci-flat four-dimensional manifold a maximally isotropic distribution $\{V_1,V_2\}$ is integrable if, and only if, the 2-form $B_{\mu\nu}=V_{1[\mu}V_{2\nu]}$ is an eigen-2-form of the Weyl operator $\textbf{C}_2$ \cite{art2}. The intent of this section is to generalize in some extent this theorem to manifolds of even dimension greater than four, more precisely it will be made a connection between the integrability of maximally isotropic distributions and some algebraic restrictions on the map $\textbf{C}_n$. Throughout this section the manifold $(M,g)$ will be assumed to have even dimension $d=2n$.

A distribution of vector fields $\{V_1,V_2,\ldots,V_p\}$ on the manifold $(M,g)$ is called isotropic or totally null when every vector field tangent to this distribution has zero norm or, equivalently, $g(V_i,V_j)=0$ for all $i,j\in\{1,2,...,p\}$. In a manifold of dimension $d=2n$ the maximum dimension that an isotropic distribution can have is $n$. If the signature is different from zero, $s\neq0$, then a maximally isotropic distribution is necessarily complex \cite{Trautman}. From now on it will always be assumed that the tangent bundle is complexified, so that the maximally isotropic distributions can exist. Now let us make the following important definitions:
\\
\\
\textbf{Definitions:} A simple $p$-form $K^{\mu_1\ldots\mu_p}=V_1^{[\mu_1}V_2^{\mu_2}\ldots V_p^{\mu_p]}$ is said to generate the distribution $\{V_1,\ldots,V_p\}$. This form will be denoted abstractly by $\textbf{K}_p=\frac{1}{p!}(V_1\wedge V_2\wedge\ldots\wedge V_p)$.
When the distribution generated by a non-zero simple $p$-form $\textbf{N}_p$ is isotropic this form is called a null $p$-form.
\\

In order to deal with totally null structures it is useful to introduce a null frame $\{e_1,\ldots,e_n,e_{n+1}=\theta^1,\ldots, e_{2n}=\theta^n \}$ on the tangent bundle, defined to be such that the only non-zero inner products are $g(e_{a'},\theta^{b'})=\frac{1}{2}\delta_{a'}^{\phantom{a}b'}$, where $a'\in\{1,2,\ldots,n\}$. In particular all the frame vectors, $e_a$, are null and the distribution $\{e_1, e_2, \ldots , e_n\}$ is maximally isotropic. In this article it will always be assumed that the indices $a,b,\ldots$ pertain to $\{1,2,\ldots,2n\}$, while the indices $a',b',\ldots$ pertain to $\{1,2,\ldots,n\}$. Given a tensor $T_{\mu\nu}$, for example, then $T_{ab}$ will denote the components of this tensor on the null frame, $T_{ab}\equiv e_a^{\;\mu}e_b^{\;\nu}T_{\mu\nu}$.
\\

In reference \cite{HigherGSisotropic2} part of the Goldberg-Sachs(GS) theorem was generalized to all dimensions greater than four and to all signatures. The content of this generalized GS theorem can be put in the following form: In a Ricci-flat manifold if the Weyl tensor is such that $C_{a'b'c'd}=0$ and generic otherwise then the maximally isotropic distribution generated by $(e_1\wedge e_2\wedge\ldots\wedge e_n)$ is locally integrable. The main purpose of this section is to express the integrability condition $C_{a'b'c'd}=0$ in terms of algebraic constraints on operator $\textbf{C}_n$. To this end it is important to define the following subbundles of $\wedge^nM$:
$$ \mathcal{A}_p \,=\,\{\,\textbf{K}_n \in \wedge^nM \;|\; e_{a'_p}\lrcorner \ldots e_{a'_2}\lrcorner e_{a'_1}\lrcorner\textbf{K}_n\,=\,0 \;\; \forall\; a'_1,\ldots,a'_p\in(1,\ldots,n) \,\}\,.  $$
Where $(V \lrcorner\textbf{K})$ is the interior product of the vector field $V$ into the differential form $\textbf{K}$. In index notation the subbundle $\mathcal{A}_p$ is the one formed by the $n$-forms $\textbf{K}_n$ such that $K_{a'_1a'_2\ldots a'_p b_1b_2\ldots b_{n-p}}=0$. For example, $\mathcal{A}_1$ has dimension one and is generated by the $n$-form $(e_1\wedge e_2\wedge\ldots\wedge e_n)$. Note that the definition of the bundles $\mathcal{A}_p$ depends on the choice of null frame. Now expanding the equation $\textbf{C}_n(\textbf{K})=\textbf{L}$ using the index notation in the null frame yields:
$$ L_{d_1d_2 \ldots d_n}\,=\,C^{ab}_{\phantom{ab}[d_1d_2}K_{d_3\ldots d_n]ab}\,=\, C^{a'b'}_{\phantom{a'b'}[d_1d_2}K_{d_3\ldots d_n]a'b'}\,+\,2 C^{a'}_{\phantom{a'}b'[d_1d_2}K_{d_3\ldots d_n]a'}^{\phantom{d_3\ldots d_n]a'}b'} \,+\,   C_{a'b'[d_1d_2}K_{d_3\ldots d_n]}^{\phantom{d_3\ldots d_n]}a'b'}$$
Taking care of the different types of indices it is possible to see that if $C_{a'b'c'd}=0$ then all subspaces $\mathcal{A}_p$ are invariant under the  action of the operator $\textbf{C}_n$. Conversely, careful calculations show that if both subbundles $\mathcal{A}_1$ and $\mathcal{A}_2$ are invariant under $\textbf{C}_n$ then  $C_{a'b'c'd}=0$ (this result is most easily seen using the notation introduced in section \ref{Sec- SimpleNot}). So we can state the following theorem and its immediate corollary:
\begin{theorem}\label{Theor. Integ.Cond.}
The following three statements are equivalent: (1) The Weyl tensor obeys to the integrability condition $C_{a'b'c'd}=0$; (2) The subbundles $\mathcal{A}_1$ and $\mathcal{A}_2$ are invariant under the action of $\textbf{C}_n$; (3) All subbundles $\mathcal{A}_p$, $p\in\{1,2,\ldots, n\}$, are invariant by the action of $\textbf{C}_n$.
\end{theorem}
\textbf{Corollary}\;\,\emph{In a Ricci-flat manifold of dimension $d=2n$ if the spaces $\mathcal{A}_1$ and $\mathcal{A}_2$ are invariant under the operator $\textbf{C}_n$ and the Weyl tensor is otherwise generic then the maximally isotropic distribution generated by $(e_1\wedge e_2\wedge\ldots\wedge e_n)$ is locally integrable.}
\\
\\
Where the main theorem of reference \cite{HigherGSisotropic2} was used to arrive at the above corollary. Since $\mathcal{A}_1$ is the one-dimensional subbundle generated by $e_1\wedge e_2\wedge \ldots \wedge e_n$, the invariance of $\mathcal{A}_1$ is equivalent to say that this null $n$-form is an eigen-form of the operator $\textbf{C}_n$. So, in particular, theorem \ref{Theor. Integ.Cond.} implies that if the integrability condition of the distribution generated by $e_1\wedge e_2\wedge \ldots \wedge e_n$ is satisfied then we have, $\textbf{C}_n(e_1\wedge e_2\wedge \ldots \wedge e_n)\propto e_1\wedge e_2\wedge \ldots \wedge e_n$. The converse of this implication is true only in four dimensions. Note also that since the bundles of self-dual and anti-self-dual $n$-forms, $S^\pm$, are invariant under $\textbf{C}_n$ then if $\mathcal{A}_p$ is invariant by this operator then the subbundles $\mathcal{A}^\pm_p\equiv \mathcal{A}_p\cap S^\pm$ are also invariant. These results strengthen the relevance and the utility of the operators $\textbf{C}_p$ introduced on section \ref{Sec- C_p}. Now let us see explicitly the application of this theorem in four and six dimensions:
\begin{quote}
 \textbf{Example($d=4$): }  In four-dimensional manifolds, $n=2$, we have  $\mathcal{A}_1=\mathcal{A}^+_1= \textrm{Span}\{e_1\wedge e_2\}$, $\mathcal{A}^+_2 =  \textrm{Span}\{ e_1\wedge e_2, (e_1\wedge e_3+e_2\wedge e_4)\}$ and $\mathcal{A}^-_2 = \textrm{Span}\{e_1\wedge e_4, (e_1\wedge e_3-e_2\wedge e_4), e_2\wedge e_3\} = S^-$. Looking at the representation of the operators $\textbf{C}^\pm$ given in reference \cite{art1} we conclude that the invariance of these spaces under the Weyl operator $\textbf{C}_n$ is equivalent to the vanishing of the Weyl scalars $\Psi_0^+=C_{1212}$ and $\Psi_1^+=C_{1213}$, this is the content of the well known Goldberg-Sachs theorem \cite{Plebanski2,art2}.
\end{quote}
\begin{quote}
 \textbf{Example($d=6$): } When the dimension of the manifold is six we have $\mathcal{A}_1=\mathcal{A}^+_1= \textrm{Span}\{e_1\wedge e_2\wedge e_3\}$, $\mathcal{A}^+_2 =  \textrm{Span}\{ e_1\wedge e_2\wedge e_3, \,e_1\wedge(e_2\wedge e_5 + e_3\wedge e_6),\, e_2\wedge(e_1\wedge e_4 +  e_3\wedge e_6),\, e_3\wedge(e_1\wedge e_4 + e_2\wedge e_5)\}$, $\mathcal{A}^-_2 = (\mathcal{A}^+_2)^\bot \cap S^-$, $\mathcal{A}^+_3 = S^+$ and  $\mathcal{A}^-_3 = (\mathcal{A}^+_1)^\bot\cap S^-$. Where $\mathcal{A}^\bot$ is the orthogonal complement of $\mathcal{A}$ with respect to the inner product defined on (\ref{Inner product}). Because of equation (\ref{C_p Autoadjunto}) it follows from these orthogonality relations that impose the invariance of $\mathcal{A}^+_1$ is equivalent to impose the invariance of $\mathcal{A}^-_3$ as well as the invariance of $\mathcal{A}^+_2$ is equivalent to the invariance of $\mathcal{A}^-_2$. The fact that the invariance of $\mathcal{A}^+_p$ is equivalent to the invariance of $\mathcal{A}^-_{n+1-p}$ and the invariance of $\mathcal{A}^-_p$ is equivalent to the invariance of $\mathcal{A}^+_{n+1-p}$, for $p\in\{1,2,\ldots,\frac{n+1}{2}\}$, seems to be a general feature of the cases in which $n$ is odd. Now using theorem \ref{Theor. Integ.Cond.} we find that the integrability condition $C_{a'b'c'd}=0$ is equivalent to the invariance of the spaces $\mathcal{A}^+_1$ and $\mathcal{A}^+_2$. This agrees perfectly with the results found on reference \cite{Spin6D} by means of spinorial calculus.
\end{quote}

Note from the first example, $d=2n=4$, that the invariance of the spaces $\mathcal{A}_1$ and $\mathcal{A}_2$ impose no restrictions on the operator $\textbf{C}^-$. This happens because in this case $\mathcal{A}^-_1=0$ and $\mathcal{A}^-_2=S^-$, so that the invariance of these spaces is always guaranteed. This phenomenon is exclusive of the four-dimensional case, in higher dimensions the invariance of $\mathcal{A}_2$ imposes restrictions on both $\textbf{C}^+$ and $\mathbf{C}^-$. This is already clear when $n$ is odd, since in these cases $\textbf{C}^+$ and $\textbf{C}^-$ carry the same degrees of freedom, one is the transpose of the other, as proved on theorem \ref{Theor. C^+andC^-}. But even if $n$ is even, besides the constraints on $\textbf{C}^+$, we have that $\textbf{C}^-$ is constrained by the integrability condition $C_{a'b'c'd}=0$, which stems from the fact that  $\dim(\mathcal{A}^-_2)=\frac{1}{2}(n+n^2)\leq \dim(S^-)=\frac{1}{2} \left(^{2n}_{\,\,n}\right)$. Thus if $n>2$ then $\mathcal{A}^-_2$ is a proper subspace of $S^-$, which implies that $\textbf{C}^-$ must admit a non-trivial invariant subspace if such integrability condition is satisfied. This remarkable difference between the well known four-dimensional case and the higher-dimensional cases will be enunciated as a lemma:
\begin{lemma}
In four dimensions the integrability condition of the distribution generated by the self-dual null $2$-form $(e_1\wedge e_2)$ imposes restrictions only on $\textbf{C}^+$, so that $\textbf{C}^-$ can be arbitrary. But in higher dimensions the integrability condition of the maximally isotropic distribution generated by the self-dual null $n$-form $(e_1\wedge e_2\wedge\ldots\wedge e_n)$ constrains both operators, $\textbf{C}^+$ and $\textbf{C}^-$.
\end{lemma}

\section{An Elegant Notation}\label{Sec- SimpleNot}

In this section it will be introduced a notation to deal with the operators $\textbf{C}_p$ that is elegant and promising. Such notation will make clear that these operators are intimately related to the integrability of distributions.

Let us denote by $\{e^a\}$ the frame of 1-forms dual to the null frame $\{e_a\}$, $e^a(e_b)=\delta^a_{\phantom{a}b}$. Then for a vanishing Ricci tensor it follows that the Cartan structure equations are:
\begin{equation*}
  \textbf{d}e^a + \omega^a_{\phantom{a}b}\wedge e^b = 0 \quad \quad \textrm{and} \quad \quad  \mathbb{C}^a_{\phantom{a}b} = \textbf{d}\omega^a_{\phantom{a}b} + \omega^a_{\phantom{a}c} \wedge \omega^c_{\phantom{c}b}\,.
 \end{equation*}
Where $\mathbb{C}^a_{\phantom{a}b}\, \equiv \, \frac{1}{2}\,C^a_{\phantom{a}bcd}\,e^{c}\wedge e^{d}$ are the curvature 2-forms when the Ricci tensor vanishes and $\omega^a_{\phantom{a}b}$ are the connection 1-forms of the Levi-Civita connection, \textit{i.e.}, 1-forms defined by the relation $\nabla_V e_b = \omega^a_{\phantom{a}b}(V) e_a$ for any vector field $V$. Taking the exterior derivative of the second Cartan equation it is simple matter to see that:
\begin{equation}\label{dC_ab}
  \textbf{d}\mathbb{C}^a_{\phantom{a}b} \,=\,  \mathbb{C}^a_{\phantom{a}c}\wedge \omega^c_{\phantom{c}b} - \omega^a_{\phantom{a}c} \wedge \mathbb{C}^c_{\phantom{c}b}\,.
\end{equation}

If $\textbf{K}_p$ is a $p$-form then we can associate to it a set of $(p-2)$-forms $\mathbb{K}^{\phantom{a}b}_{a}$, with $a,b\in\{1,2,\ldots,d\}$, defined by:
\begin{equation*}\label{Kab_def.}
  \mathbb{K}^{\phantom{a}b}_{a}\, \equiv \,\frac{2}{p!}\, K^{\phantom{a}b}_{a\phantom{ab}c_1c_2\ldots c_{p-2}}\, e^{c_1}\wedge e^{c_2}\wedge \ldots \wedge e^{c_{p-2}}\,.
\end{equation*}
Now using the just defined objects note that
$$\mathbb{C}^a_{\phantom{a}b}\wedge \mathbb{K}^{\phantom{a}b}_{a} \,=\, \frac{1}{p!} C^a_{\phantom{a}bc_1c_2} K^{\phantom{a}b}_{a\phantom{b}c_3c_4\ldots c_{p}}\, e^{c_1}\wedge e^{c_2}\wedge \ldots \wedge e^{c_{p}} = \frac{1}{p!} C^{ab}_{\phantom{ab}[c_1c_2} K_{c_3c_4\ldots c_{p}]ab}\, e^{c_1}\wedge e^{c_2}\wedge \ldots \wedge e^{c_{p}} \,.$$
By definition of $\textbf{C}_p$, equation (\ref{C_p}), the last relation implies the following elegant form to express the action of this operator:
\begin{equation}\label{C_p elegant}
  \textbf{C}_p(\textbf{K}_p) \,=\,  \mathbb{C}^a_{\phantom{a}b}\wedge \mathbb{K}^{\phantom{a}b}_{a}\,.
\end{equation}

Now we are ready to see one more hint that the operator $\textbf{C}_p$ is connected to the integrability of distributions. First let us define the $(p-1)$-form $\textbf{D}\mathbb{K}^{\phantom{a}b}_{a}$ by $\textbf{D}\mathbb{K}^{\phantom{a}b}_{a}\equiv \textbf{d}\mathbb{K}^{\phantom{a}b}_{a} +  \omega^b_{\phantom{b}c}\wedge \mathbb{K}_{a}^{\phantom{a}c} - \omega^c_{\phantom{c}a}\wedge \mathbb{K}_{c}^{\phantom{c}b}$. Then using the definition of $\mathbb{K}^{\phantom{a}b}_{a}$ and using the metric to lower (or raise) the indices it easily follows that $\textbf{K}_p=\frac{1}{2}\mathbb{K}_{ab}\wedge e^{a}\wedge e^{b}$. Taking the exterior derivative of this relation and using the first structure equation we find $\textbf{d}\textbf{K}_p=\frac{1}{2}   e^{a}\wedge e^{b}\wedge\textbf{D}\mathbb{K}_{ab}$, where $\textbf{D}\mathbb{K}_{ab}\equiv g_{bc}\textbf{D}\mathbb{K}^{\phantom{a}c}_{a}$. Also taking the exterior derivative of equation (\ref{C_p elegant}) and using equation (\ref{dC_ab}) it follows that $\textbf{d}\left[\textbf{C}_p(\textbf{K}_p) \right] =\mathbb{C}^{ab}\wedge \textbf{D}\mathbb{K}_{ab}$. Let us summarize these important relations:
\begin{equation}\label{dK dC(k)}
  \textbf{d}\textbf{K}_p \,=\,\frac{1}{2}\,   e^{a}\wedge e^{b}\wedge \textbf{D}\mathbb{K}_{ab} \quad\quad ; \quad\quad \textbf{d}\left[\textbf{C}_p(\textbf{K}_p) \right] \,=\, \mathbb{C}^{ab}\wedge \textbf{D}\mathbb{K}_{ab}\,.
\end{equation}

Suppose that a simple $p$-form $\textbf{L}_p$ is in the kernel of $\textbf{C}_p$, $\textbf{C}_p(\textbf{L}_p)=0$, and that $C_{ab}^{\phantom{ab}cd}\textbf{D}\mathbb{L}_{cd}=\lambda\textbf{D}\mathbb{L}_{ab}$, with $\lambda\neq 0$. Then equation (\ref{dK dC(k)}) implies that $0=\mathbb{C}^{ab}\wedge \textbf{D}\mathbb{L}_{ab}= \frac{1}{2}C^{ab}_{\phantom{ab}cd}e^c\wedge e^d\wedge \textbf{D}\mathbb{L}_{ab}=\frac{\lambda}{2} e^c\wedge e^d \wedge \textbf{D}\mathbb{L}_{cd}$. Introducing this in the first relation of (\ref{dK dC(k)}) we see that the simple form $\textbf{L}_p$ is closed, $\textbf{d}\textbf{L}_p=0$.  This, in turn, implies that the distribution annihilated by $\textbf{L}_p$ is integrable (more about this on section \ref{Sec- Mariot-Rbinson}).

Now if $\textbf{K}_p$ is a $p$-form such that $\textbf{D}\mathbb{K}_a^{\phantom{a}b}=\kappa\wedge\mathbb{K}_a^{\phantom{a}b}$ for some 1-form $\kappa$, then equation (\ref{dK dC(k)}) implies that $\textbf{d}\textbf{K}_p = \kappa\wedge\textbf{K}_p$ and $\textbf{d}\left[\textbf{C}_p(\textbf{K}_p) \right] =\kappa\wedge\textbf{C}_p(\textbf{K}_p) $. So if $\textbf{K}_p$ is a simple $p$-form then the first equation implies that the distribution annihilated by $\textbf{K}_p$ is integrable and, analogously, if $\textbf{C}_p(\textbf{K}_p)$ is a simple $p$-form then second equality implies that the distribution annihilated by $\textbf{C}_p(\textbf{K}_p)$ is integrable. Thus connecting the operator $\textbf{C}_p$ with one more integrability property.

These are just simple results that follow from equation (\ref{dK dC(k)}), but the elegance of the expressions obtained in this section together with the results of appendix \ref{Appendix-SimpleForms} seems to denounce that the notation introduced here can be exploited to find extensions of the Goldberg-Sachs theorem. Hopefully this will be done elsewhere.

\section{Other Classification Methods}\label{Sec- OtherClassif.}

The intent of the present section is to briefly digress about other forms of classifying the curvature of the tangent bundle and, when possible, relate in some way these classification schemes to the classification for the Weyl tensor defined in the preceding sections.
\newline

The CMPP classification is the most widespread and successful scheme of classification for the Weyl tensor in higher-dimensional general relativity \cite{CMPP,Coley}. Such scheme makes use of a one-dimensional subgroup of the orthogonal transformations on the tangent bundle to split the Weyl tensor as a sum of terms with different boost weights. This subgroup is determined once a real null vector field, $l$, is chosen. If it is possible to choose $l$ in order to annihilate the components of the Weyl tensor with boost weight 2 then $l$ is said to be a Weyl aligned null direction(WAND). Further, if this null direction is such that all Weyl tensor components with boost weights 2 and 1 vanish then $l$ is called a multiple WAND.

Let $e_1$ and $\theta^1$ be real vector fields such that $\{e_1,e_2,\theta^1,\theta^2\}$ is a null frame in a 4-dimensional Lorentzian manifold. It is a established result that the null 2-form $e_1\wedge e_2$ is an eigen-2-form of operator $\textbf{C}_2$ if, and only if, the vector field $e_1$ is a multiple WAND \cite{art2}. Simple calculations also reveal that in 5-dimensional spacetimes if $\textbf{C}_2$ admits a null eigen-2-form then the real null direction tangent to the isotropic plane generated by this 2-form will be a multiple WAND. But the converse of this result is not true anymore, \textit{i.e.}, in 5 dimensions the existence of a multiple WAND does not imply the existence of a null eigen-2-form for $\textbf{C}_2$. Now it is natural wondering whether similar relations are verified in higher-dimensional Lorentzian manifolds. Using the notation introduced in the last section it can be proved that if $\boldsymbol{N}=e_1\wedge e_2\wedge\ldots\wedge e_n$ is a null $n$-form in a Lorentzian manifold of dimension $d=2n+\epsilon$, with $\epsilon$ equal to 0 or 1 and $n\geq3$, such that $e_1$ is a real vector field then the equation $\textbf{C}_n\left(\boldsymbol{N}\right)\propto\boldsymbol{N}$ does not imply that $e_1$ is a WAND, let alone a multiple WAND. Conversely, if $e_1$ is a multiple WAND then generally it is not true that $\textbf{C}_n\left(\boldsymbol{N}\right)\propto\boldsymbol{N}$. Particularly, these results can be easily verified in 6 dimensions by means of the spinorial formalism of reference \cite{Spin6D}.

Although it was argued in the last paragraph that there is no immediate relation between a WAND and an eigen-$p$-form of the operator $\textbf{C}_p$ in higher-dimensional spacetimes, these Weyl operators are still valuable for the CMPP classification. Just as reference \cite{BivectHighDim} has used the well known bivector operator $\textbf{C}_2$ to refine this classification, the other operators $\textbf{C}_p$ can analogously be used to further refine the CMPP classification. For this task the general aspects of the operator method for classifying tensors in higher-dimensional manifolds described in \cite{BivectHighDim,Hervik-Coley} could be helpful.
\newline

Another useful classification for the Weyl tensor is the Taghavi-Chabert classification \cite{HigherGSisotropic2}. As well as CMPP classification is associated to a null direction, the Taghavi-Chabert classification needs a maximally isotropic distribution to be chosen. According to this classification the Weyl tensor of an even-dimensional manifold ($d=2n\geq6$) can be of the following types, from the most general to the most special: $\mathcal{C}^{-2}$, $\mathcal{C}^{-1}$, $\mathcal{C}^{0}$, $\mathcal{C}^{1}$, $\mathcal{C}^{2}$ and $\mathcal{C}^{3}$. Using the formalism of section \ref{Sec- MaximallyIstrop} it can be proved that if the subbundle $\mathcal{A}_1 \subset \wedge^nM$ is invariant under the operator $\textbf{C}_n$ then the Weyl tensor is more special than type $\mathcal{C}^{-2}$, but the converse is not true. Analogously, theorem \ref{Theor. Integ.Cond.} implies that the Weyl tensor is type $\mathcal{C}^{0}$, or more special, if, and only if, both subbundles $\mathcal{A}_1$ and $\mathcal{A}_2$ are invariant under $\textbf{C}_n$. It would be interesting, and hopefully valuable, to use the algebraic classification introduced in section \ref{Sec- C_p} to refine the Taghavi-Chabert classification, \textit{i.e.}, investigate how each of the types of the latter classification constrain the refined Segre types of the operators $\textbf{C}_p$.
\newline

Finally, a geometric well known form of classifying the curvature of the tangent bundle is using the holonomy group associated to the Levi-Civita connection. In a connected manifold the holonomy is the same at all points \cite{Nakahara}, thus providing a global classification, whereas previously seen classifications are local\footnote{For example, in \cite{Pravda05} it was shown that the CMPP type on the horizon of a 5-dimensional black ring spacetime is different from the CMPP type outside the horizon.}. Particularly, in four-dimensional Lorentzian manifolds the holonomy classification has been widely studied and compared with Petrov classification \cite{Hall-Schell}, in which case equation (\ref{Commutation}), with $d=4$ and $p=2$, is behind some simplifications of this analysis. In higher-dimensional Lorentzian manifolds important developments have recently been achieved on this subject, see for example \cite{Galaev,Leistner} and references therein.

Purely gravitational solutions of supergravity can be put in one-to-one correspondence with Ricci-flat Lorentzian manifolds admitting a covariantly constant spinor \cite{M-waves}. By means of the Ambrose-Singer theorem it follows that the integrability condition for a constant spinor imposes restrictions on the holonomy\footnote{Generally the existence of a covariantly constant tensor on the manifold imposes restrictions on the holonomy. For instance, K\"{a}hler manifolds are characterized by the existence of a covariantly constant complex structure and because of this their holonomy are subgroups of $U(n)\subset SO(2n)$, where $n$ is the complex dimension of the manifold.} \cite{M-waves}. Therefore the holonomy group is of fundamental importance to supergravity and, consequently, to string theory. Holonomy classification also provides powerful tools to find solutions to Einstein's equation \cite{Galaev}. Moreover, holonomy is of major relevance to loop quantum gravity \cite{LoopQG}, although in a broader sense, since the holonomy may be associated to the curvature of a gauge field.

\section{Generalized Mariot-Robinson Theorem}\label{Sec- Mariot-Rbinson}
The Mariot-Robinson theorem states that in four dimensions a null 2-form, $\textbf{F}'$, generates a locally integrable distribution of planes if, and only if, there exists some function $h\neq0$ such that $\textbf{F}=h\textbf{F}'$ satisfies Maxwell's equations, $\textbf{d}\textbf{F}=0$ and $\textbf{d}( \widetilde{\textbf{F}})=0$ \cite{Robinson,McIntoshII}. Where $ \widetilde{\textbf{F}}$ is the Hodge dual of $\textbf{F}$ and $\textbf{d}$ is the exterior derivative operator. In the present section this theorem will be generalized in a natural way to all dimensions.

Following similar steps to the ones taken in \cite{McIntoshII} let $\{\omega^1, \ldots , \omega^p\}$ be linearly independent 1-forms of a $d$-dimensional manifold such that $\textbf{d}(h \omega^1\wedge\ldots\wedge\omega^p)=0$ for some function $h\neq0$. Expanding this equation and taking the wedge product with $\omega^i$ it is immediate to see that $(\textbf{d} \omega^i)\wedge\omega^1\wedge\ldots\wedge\omega^p=0$. It is a standard result of differential geometry that this implies the local integrability of the distribution of vector fields annihilated by $\{\omega^1, \ldots, \omega^p\}$ \cite{Frankel}. Conversely, if the distribution annihilated by $\{\omega^i\}$ is integrable then it follows that there exist $p^2$ functions $f^i_{\phantom{i}j}$  and $p$  coordinates $x^j$ such that $\omega^i = f^i_{\phantom{i}j}\textbf{d}x^j$, with $\det(f^i_{\phantom{i}j})=f\neq0$. Then it is easy to see that $\textbf{d}(\frac{1}{f}\omega^1\wedge\ldots\wedge\omega^p) = \textbf{d}(\textbf{d}x^1\wedge \ldots\wedge \textbf{d}x^p) = 0$. Thus we arrived at the following result: \emph{The distribution of vector fields annihilated by $\{\omega^1, \ldots, \omega^p\}$ is integrable if, and only if, there exists some function $h\neq0$ such that} $\textbf{d}(h \omega^1\wedge\ldots\wedge\omega^p)=0$.

Now let us suppose that the dimension of the manifold is even, $d=2n$. As defined in section \ref{Sec- MaximallyIstrop}, an $n$-form $\textbf{N}_n$ is called null if it can be written as $N^{\mu_1\mu_2\ldots\mu_n}= V_1^{[\mu_2}V_1^{\mu_2} \ldots V_n^{\mu_n]}$ where the vector fields $\{V_1,\ldots,V_n\}$ form an  isotropic distribution of dimension $n$. Such distribution is said to be generated by $\textbf{N}_n=V_1\wedge\ldots\wedge V_n$ and in this case, since the isotropic distribution is maximal, it is also the distribution annihilated by $\textbf{N}_n$. As proved in appendix \ref{Appendix-SimpleForms} null $n$-forms must be self-dual or anti-self-dual, meaning that $\textbf{E}_n(\textbf{N}_n)\equiv \widetilde{\textbf{N}}_n=\pm\epsilon\,\textbf{N}_n$, where $\epsilon$ is equal to $1$ or $i$ depending on the signature of the manifold. So  for a null $n$-form the equation $\textbf{d}\textbf{N}_n=0$ is equivalent to the equation $\textbf{d} \widetilde{\textbf{N}}_n=0$. Thus using the results of this and of the last paragraph we arrive at the following theorem:
\begin{theorem}
In a space of dimension $d=2n$, the null n-form $\textbf{N}_n'$ generates an integrable maximally isotropic distribution  if, and only if, there exists some function $h\neq0$ such that $\textbf{N}_n=h\textbf{N}_n'$ obeys to the equations $\textbf{d}\textbf{N}_n=0$ and $\textbf{d} \widetilde{\textbf{N}}_n =0$.
\end{theorem}

It is important to note that, contrary to the Goldberg-Sachs theorem, no condition on the Ricci tensor was assumed. This theorem was proved before on reference \cite{Kerr-Robinson}, but there the proof uses spinor and twistor calculus and is less concise.

\section{Optical Matrix and the Forms that are Closed and Co-closed}\label{Sec- OpticalMatrix}

Let $\{l,n,m_2,m_3,\ldots,m_{(d-1)}\}$ be a semi-null frame of the tangent bundle such that the only non-zero inner products are $g(l,n)=1$ and $g(m_i,m_j)=\delta_{ij}$, in particular the vectors $l$ and $n$ are null. Then denoting by $\nabla$ the Levi-Civita connection let us define the following quantities:
\begin{equation}\label{Optical}
 M_0\,=\, l^\nu n^\mu \, \nabla_\nu l_\mu\ \;\;\;;\;\; M_i\,=\, l^\nu m_i^\mu \, \nabla_\nu l_\mu\ \;\;\;;\;\; M_{ij}\,=\, m_j^\nu m_i^\mu \, \nabla_\nu l_\mu\ \,.
\end{equation}
Note that in Lorentzian signature the vectors of the frame $\{l,n,m_i\}$ can be chosen to be real, so that the quantities $M_0$, $M_i$ and $M_{ij}$ are real in this case. The $(d-2)\times(d-2)$ matrix $M_{ij}$ is called the optical matrix and is usually decomposed as the sum of a symmetric trace-less matrix $\sigma_{ij}$, the shear, a skew-symmetric matrix $A_{ij}$, the twist, and a term proportional to the identity matrix $\theta\delta_{ij}$, the expansion.
\begin{equation}\label{Optical Decomp.}
  M_{ij}\,=\,  \sigma_{ij} \,+\, A_{ij} \,+\, \theta\delta_{ij} \;;\quad  \theta=\frac{1}{d-2}\delta^{ij}M_{ij}\;;\quad \sigma_{ij}=M_{(ij)}-\theta\delta_{ij}\;;\quad A_{ij}=M_{[ij]}
\end{equation}

It is easy matter to establish that the null congruence generated by $l$ is geodesic if, and only if, $M_i=0$ and the parametrization is affine when $M_0=0$. Also, straightforward calculations show that the congruence generated by $l$ is hyper-surface-orthogonal, $l_{[\alpha} \nabla_\mu l_{\nu]} =0$, if, and only if, $M_i$ and $A_{ij}$ both vanish.

Now let $\textbf{K}_p$ be a non-zero $p$-form obeying to the equations $\textbf{dK}_p =0$ and $\textbf{d}\widetilde{\textbf{K}}_p =0$ and such that $l$ is a multiple aligned null direction \cite{CMPP,Coley} of $\textbf{K}_p$. Using index notation these constraints are respectively given by:
\begin{equation}\label{Null p-form}
  \nabla_{[\alpha}\, K_{\mu_1\mu_2 \ldots \mu_p]} =0 \;\;;\quad  \nabla^{\alpha}\, K_{\alpha\mu_2 \ldots \mu_p}=0  \;\;;\quad  K^{\mu_1\mu_2 \ldots \mu_p}= p!f_{j_2j_3\ldots j_p}\, l^{[\mu_1}m_{j_2}^{\mu_2}m_{j_3}^{\mu_3}\ldots m_{j_p}^{\mu_p]}\,.
\end{equation}
Where $f_{j_2j_3\ldots j_p}$ is completely antisymmetric and in the last relation it is being assumed a sum over the indices $j_2j_3\ldots j_p$. Let us prove that the existence of a non-zero $p$-form that satisfies equation (\ref{Null p-form}) imposes restrictions on the optical matrix as well as on $M_i$. Developing the equation $0=(\nabla^{\alpha} K_{\alpha\mu_2 \ldots \mu_p})l^{\mu_2}= -K_{\alpha\mu_2 \ldots \mu_p} \nabla^{\alpha} l^{\mu_2} $  the following results can be obtained:
\begin{equation}\label{M_i,A_ij}
   M_i\, f_{ij_3\ldots j_p} \,=\,0 \;\;;\quad  A_{ij}f_{ij k_4\ldots k_p}\,=\,0\,.
\end{equation}
Analogously, expanding the equation $0=\nabla_{[\alpha}\, K_{\mu_1\mu_2 \ldots \mu_p]} l^\alpha m_{j_1}^{\phantom{j_1} \mu_1}\ldots m_{j_p}^{\phantom{j_p} \mu_p}$ we get:
\begin{equation*}\label{M_[i f]}
  M_{[j_1}\,f_{j_2 \ldots j_p]} \,=\, 0\,.
\end{equation*}
Note that contracting the above equation with $M_{j_1}$ and using equation (\ref{M_i,A_ij}) we get that $M_jM_j=0$. In the particular case of Lorentzian signature the frame $\{l,n,m_i\}$ can be real, then $M_i$ is real and the relation $M_jM_j=0$ implies that $M_j$ vanish, \textit{i.e.}, the null congruence generated by $l$ is geodesic.
On the same vein, after workout the equality $0=(\nabla^{\alpha}\, K_{\alpha\mu_2 \ldots \mu_p})m_{j_2}^{\phantom{j_2} \mu_2}\ldots m_{j_p}^{\phantom{j_p} \mu_p}$ it follows that:
\begin{equation}\label{K Nabla(m)}
  K_{\alpha\mu_2 \ldots \mu_p} \nabla^\alpha (m_{j_2}^{\phantom{j_2} \mu_2}\ldots m_{j_p}^{\phantom{j_p} \mu_p})\,=\, (p-1)!\,\, l^\alpha \nabla_\alpha f_{j_2\ldots j_p} \,+\, (p-1)!\, f_{j_2\ldots j_p} \nabla^\alpha l_\alpha\,.
\end{equation}
Now expanding the relation $0=\left(\nabla_{[\alpha}\, K_{\mu_1\mu_2 \ldots \mu_p]}\right)l^\alpha n^{\mu_1}m_{j_2}^{\phantom{j_2} \mu_2}\ldots m_{j_p}^{\phantom{j_p} \mu_p}$, using the identity $\nabla^\alpha l_\alpha = M_0 + (d-2)\theta$ and the equation (\ref{K Nabla(m)}) it follows, after some algebra, that:
\begin{equation*}\label{Shear f}
  2(p-1)\,f_{i[j_3 \ldots j_p} \, \sigma_{j_2] i} \;=\; (d-2p)\,\theta\, f_{j_2 \ldots j_p}\,.
\end{equation*}
These results are summarized by the following theorem:

\begin{theorem}\label{Theor. Optical Scalars}
If  $K^{\mu_1\mu_2 \ldots \mu_p}= p!f_{j_2\ldots j_p}\, l^{[\mu_1}m_{j_2}^{\mu_2}\ldots m_{j_p}^{\mu_p]}$ is a non-zero $p$-form such that $\textbf{dK}_p =0$ and $\textbf{d}\widetilde{\textbf{K}}_p =0$ then it follows that:
\begin{itemize}
  \item $M_i\, f_{ij_3\ldots j_p} \,=\,0$
  \item $ M_{[j_1}\,f_{j_2 \ldots j_p]} \,=\, 0$
  \item $2(p-1)\,f_{i[j_3 \ldots j_p} \, \sigma_{j_2] i} \;=\; (d-2p)\,\theta\, f_{j_2 \ldots j_p}$
  \item $M_iM_i\,=\,0$
  \item $A_{ij}f_{ij k_4\ldots k_p}\,=\,0$
\end{itemize}
In the particular case of a real frame $\{l,n,m_i\}$ (Lorentzian signature) it follows from the fourth point above that $M_i=0$, \textit{i.e.}, the vector field $l$ is geodesic.
\end{theorem}

Particularly, if the signature is Lorentzian and the dimension is four, $d=4$, then the case $p=2$ of the above theorem implies that if a real null bivector $\textbf{F}$ obeys to the source-free Maxwell's equations then the real null direction $l$ such that $F_{\mu\nu}l^\nu=0$ generates a congruence that is geodesic and shear-free (shear-free meaning that $\sigma_{ij}=0$), a classical result first obtained a long time ago by Ivor Robinson \cite{Robinson}. In the Lorentzian signature the case $p=2$ of the above theorem was obtained before on reference \cite{M. Ortaggio-Robinson-Trautman}. Also in the Lorentzian case these results can be easily obtained using the generalized GHP formalism of reference \cite{GHP}\footnote{Thanks to Harvey S. Reall for gently pointing out this reference.}. More precisely the first three points of the above theorem can be found in the proof of Lemma 3 of reference \cite{GHP}, whereas the last two points of the above theorem are not explicit in \cite{GHP}, although can be easily derived using the GHP formalism. But it is worth remarking that there is an important difference between the Lorentzian and the non-Lorentzian signatures: while in the former case the equation $M_iM_i=0$ implies that the null vector field $l$ is geodesic in the latter this is not true in general. Note also that in the Euclidean case we have that $n$ is the complex conjugate of $l$, therefore in this signature any constraint on the optical matrix of $l$ imposes an analogous constraint on the optical matrix of $n$.

%
%
%

\section{A Generalized Version of the Goldberg-Sachs Theorem}\label{Sec- GSgeneralized}
Using the previous results it will be proved in this section a generalized version of the Goldberg-Sachs theorem that is valid in any even dimension and that makes no assumption about the Ricci tensor, contrary to the usual generalizations of such theorem. The theorem presented here states that if there exists an integrable maximally isotropic distribution on a manifold of even dimension then the shear matrix of the null directions tangent to such distribution must be constrained.

Before proceeding let us introduce some notation that will be used in what follows. In a manifold of dimension $d=2n$ let $\{e_1,e_2,\ldots,e_n\}$ be a maximally isotropic distribution of vector fields. We can complete this distribution to form a null frame $\{e_1,\ldots,e_n,e_{n+1}=\theta^1,\ldots, e_{2n}=\theta^n \}$ such that the only non-zero inner products are $g(e_{a'},\theta^{b'})=\frac{1}{2}\delta_{a'}^{\phantom{a}b'}$. From this frame we can construct a semi-null frame, like the one used in the latter section, $\{l,n,m_2,m_3,\ldots,m_{(d-1)}\}$ defined by:
\begin{equation*}\label{Frame l,n,m}
 l=e_1\;\;;\;\; n=2\theta^1 \;\;;\;\; m_j=(e_j+\theta^j) \;\;;\;\;  m_{j+n-1}=-i(e_j-\theta^j)\;,\;\;\;\textrm{where}\;\; j\,\in\,\{2,3,\ldots,n\}\,.
\end{equation*}
Note that in general the choice of the vector $e_1$ to be $l$ is arbitrary, we could have chosen any vector on the distribution $\{e_1,\ldots,e_n\}$ to take this place. In the special case of Lorentzian signature there is a privileged choice, since in this case a maximally isotropic subspace admits just one real direction \cite{Trautman}, in the present section whenever the signature is Lorentzian the real direction will be assumed to be tangent to the vector field $e_{1}=l$.

Once defined such semi-null frame we can define the optical matrix, the shear, the twist and the expansion associated to the distribution $\{e_1,\ldots,e_n\}$ to be just as defined in equations (\ref{Optical}) and (\ref{Optical Decomp.}).  Also from this maximally isotropic distribution we can construct the null $n$-form $\textbf{N}_n=e_1\wedge e_2\wedge \ldots \wedge e_n$.  In the semi-null frame just defined this $n$-form has the following expansion:
\begin{equation}\label{Nexpansion}
N^{\mu_1\mu_2\ldots\mu_n}\,\equiv\, n!\, e_1^{\,[\mu_1}\ldots e_n^{\,\mu_n]}\,=\, n!\, \widehat{f}_{j_2j_3\ldots j_n} l^{[\mu_1}m_{j_2}^{\,\mu_2}\ldots m_{j_n}^{\,\mu_n]}.
\end{equation}
With $\widehat{f}_{j_2j_3\ldots j_n}=\widehat{f}_{[j_2j_3\ldots j_n]}$ different from zero only when each index $j_r$ is equal to $r$ or to $r+n-1$ (and permutations of this), in which case  $\widehat{f}_{j_2j_3\ldots j_n}= 2^{1-n} \, i^q$ where $q$ is the number of indices $j_r$ that are equal to $r+n-1$. For example, in six dimensions, $n=3$, the non-zero components of $\widehat{f}_{j_2j_3}$ are:
\begin{equation}\label{f Matrix6D}
   \widehat{f}_{23}=-\widehat{f}_{32}=\frac{1}{4} \;\;;\;\; \widehat{f}_{25}=-\widehat{f}_{52}=\frac{i}{4} \;\;;\;\; \widehat{f}_{43}=-\widehat{f}_{34}=\frac{i}{4}  \;\;;\;\; \widehat{f}_{45}=-\widehat{f}_{54}=-\frac{1}{4}\,.
\end{equation}


Now we are able to prove the main result of this section. Suppose that the maximally isotropic distribution $\{e_1,\ldots,e_n\}$ is integrable. So from what was seen in section \ref{Sec- Mariot-Rbinson} there exists some function $h\neq0$ such that $\textbf{d}(h\textbf{N}_n) = 0$ and $\textbf{ d}(h\widetilde{\textbf{N}}_n) = 0$, where $\textbf{N}_n=e_1\wedge e_2\wedge \ldots \wedge e_n$. Thus using theorem \ref{Theor. Optical Scalars} and equation (\ref{Nexpansion}) it follows that $h\widehat{f}_{i[j_3 \ldots j_n} \, \sigma_{j_2] i} = 0$, which leads to the following theorem:

\begin{theorem}[Generalized GS]\label{Theor. Gen.GS}
In a manifold of even dimension $d=2n$ if the maximally isotropic distribution  $\{e_1,\ldots,e_n\}$ is integrable then the shear matrix associated to this distribution, $\sigma_{ij}$, is constrained by the following equation:
$$ \widehat{f}_{i[j_3j_4 \ldots j_n} \, \sigma_{j_2] i} \;=\; 0\,.$$
Also, the twist matrix, $A_{ij}$, and the scalars $M_i$ are constrained by the following relations:
$$A_{ij}\widehat{f}_{ij k_4\ldots k_n}\,=\,0\quad;\quad M_i\, \widehat{f}_{ij_3\ldots j_n} =0 \quad;\quad M_{[j_1}\,\widehat{f}_{j_2 \ldots j_n]} = 0 \quad;\quad M_iM_i=0.$$
Particularly, from this last relation, $M_iM_i=0$, it follows that $e_1=l$ must be geodesic in the Lorentzian signature.
\end{theorem}

It is worth noting that in this theorem no condition was imposed on the Ricci tensor. It is also relevant to mention that in  the appendix C of reference \cite{GS-Ortaggio12} the integrability of a maximally isotropic distribution is expressed in terms of the Ricci rotation coefficients of a null frame. Now let us see an example of the use of theorem \ref{Theor. Gen.GS}.

\begin{quote}
  \textbf{Example: } In six dimensions, $d=6$, if a maximally isotropic distribution $\{e_1,e_2,e_3\}$ is integrable then it follows from theorem \ref{Theor. Gen.GS} that $\widehat{f}_{i[j}\sigma_{k]i}=0$, \textit{i.e.}, $\widehat{f}_{ij}\sigma_{ki}= \widehat{f}_{ik}\sigma_{ji}$. Denoting by $\boldsymbol{f}$ the matrix $\widehat{f}_{ij}$ and by $\boldsymbol{\sigma}$ the matrix $\sigma_{ij}$ then since $\boldsymbol{f}$ is skew-symmetric and $\boldsymbol{\sigma}$ is symmetric it follows that this equation is equivalent to $\boldsymbol{f}\boldsymbol{\sigma}= -\boldsymbol{\sigma}\boldsymbol{f}$. From equation (\ref{f Matrix6D}) it follows that we can write:
$$\boldsymbol{f}\,=\,\frac{1}{4}\left[
  \begin{array}{cccc}
    0 & 1 & 0 & i \\
    -1 & 0 & -i & 0 \\
    0 & i & 0 & -1 \\
    -i & 0 & 1 & 0 \\
  \end{array}
\right]\,.$$

Then using the fact that the matrix $\boldsymbol{\sigma}$ is symmetric and imposing that it anti-commutes with $\boldsymbol{f}$ we find that the shear matrix must be of the following form:
$$\boldsymbol{\sigma}\,=\,\left[
               \begin{array}{cccc}
                 \Delta & \Theta & 0 & \Phi \\
                 \Theta & -\Delta & -\Phi & 0 \\
                 0 & -\Phi & \Delta & \Theta \\
                 \Phi & 0 & \Theta & -\Delta \\
               \end{array}
             \right]\,.
 $$
Where $\Delta$, $\Theta$ and $\Phi$ are arbitrary functions. Calculating the eigenvalues of $\boldsymbol{\sigma}$ we find $\pm\sqrt{\Delta^2+\Theta^2+\Phi^2}$, so the shear must have two degenerate eigenvalues. It must be stressed that by orthonormal transformations of the semi-null frame it is possible to simplify further the shear matrix. For example, in the Lorentzian signature in addition of being traceless and symmetric $\boldsymbol{\sigma}$ is also real, so that this matrix admits, in a suitable basis, the form $\diag(\lambda,\lambda,-\lambda,-\lambda)$ with $\lambda=\sqrt{\Delta^2+\Theta^2+\Phi^2}$. This result is perfectly compatible with the six-dimensional calculations, in Lorentzian signature, found in appendix C of \cite{GS-Ortaggio12}.
\end{quote}

Despite being called here a generalized version of the Goldberg-Sachs theorem the theorem just presented carries little resemblance with the usual version of the GS theorem. The main differences being that in theorem \ref{Theor. Gen.GS} no mention is made to the Weyl tensor and the Ricci tensor is not assumed to be constrained. In order to understand why it is adequate to call this theorem a generalized version of the GS theorem it is necessary to review some previous results on the literature, this will be done in the next paragraph.

The first apparition of the Goldberg-Sachs theorem was in reference \cite{Goldberg-Sachs}, where it was proved that a Ricci-flat four-dimensional Lorentzian manifold has an algebraically special Weyl tensor if, and only if, it admits a null congruence that is geodesic and shear-free($\sigma_{ij}=0$). Later this result has been put in a conformally invariant form by relaxing the constraint on the Ricci tensor \cite{GS-CottonYork}, particularly the GS theorem has been proved to be valid for Einstein manifolds. A decade after this the GS theorem was generalized to be valid in four-dimensional manifolds of all signatures in \cite{Plebanski2}, where it was proved that the concept of geodesic and shear-free should be substituted in the other signatures by the requirement of local integrability of totally null 2-surfaces (see also \cite{Robinson Manifolds}). Indeed, the leafs of integrable maximally isotropic distributions can be proved to be totally geodesic \cite{Plebanski2,Mason-Chabert-KillingYano}, so that in the Lorentzian case the intersection of an integrable maximally isotropic distribution with its complex conjugate yields a null congruence that is geodesic and in four dimensions is shear-free.

Now we are in position to conclude that the theorem presented in this section is deeply connected to the original version of the GS theorem. From the reference \cite{Plebanski2} it follows, in particular, that in a Ricci-flat 4-dimensional manifold if the Weyl tensor is algebraically special then the manifold admits an integrable maximally isotropic distribution of vector fields. So because of theorem \ref{Theor. Gen.GS} this implies that in four dimensions if the Weyl tensor is algebraically special then $\widehat{f}_i\sigma_{ji}=0$. Since in the Lorentzian signature the $2\times2$ matrix $\sigma_{ij}$ is real, symmetric and traceless this last equation implies that $\sigma_{ji}=0$, thus arriving at the usual form of the GS theorem. The converse of theorem \ref{Theor. Gen.GS} although valid in four dimensions probably is not valid in higher dimensions, so that this theorem captures only one direction of the original GS theorem.

A connection between algebraically special Weyl tensors and restrictions on the shear matrix can also be easily made in higher dimensions using previous results. Combining theorem \ref{Theor. Gen.GS} with the corollary of theorem \ref{Theor. Integ.Cond.} we immediately arrive at the following result:
\\
\\
\textbf{Corollary}\;\,\emph{In a Ricci-flat manifold of even dimension $d=2n$ if the subbundles $\mathcal{A}_1$ and $\mathcal{A}_2$ are invariant under the operator $\textbf{C}_n$ and this operator is generic otherwise then the shear matrix of the maximally isotropic distribution $\{e_1,\ldots,e_n\}$ is constrained by the following equation:}
$$ \widehat{f}_{i[j_3j_4 \ldots j_n} \, \sigma_{j_2] i} \;=\; 0\,.$$
\emph{ Where $\widehat{f}_{j_2j_3 \ldots j_n}$ was defined below equation (\ref{Nexpansion}) and the subbundles $\mathcal{A}_p$ were defined in section \ref{Sec- MaximallyIstrop}. In the particular case of Lorentzian signature the vector field $e_1$ is geodesic.}\\
\\
Also, although not explicitly stated, the twist matrix and the scalars $M_i$ are also constrained if the hypotheses of this corollary are satisfied (see theorem \ref{Theor. Gen.GS}). This corollary is the reason of why the theorem \ref{Theor. Gen.GS} was called here a generalized Goldberg-Sachs theorem.

\section{Conclusions and Perspectives}
In this article it was investigated several aspects of the Weyl tensor classification and its relation with the integrability of vector distributions, specially the maximally isotropic ones. At first the Weyl operators $\textbf{C}_p$ were introduced and its properties investigated. These operators are non-trivial generalizations of the already known bivector map provided by the Weyl operator, here represented by $\textbf{C}_2$, that can be used to classify the Weyl tensor. It was shown that such operators are self-adjoint with respect to the natural inner product on the space of forms and that in the Euclidean signature they can diagonalized, providing a simple form to classify the Weyl tensor.

In even dimensions, $d=2n$, the operator $\textbf{C}_n$ has the special property of preserving the space of (anti-)self-dual $n$-forms, so that we can write $\textbf{C}_n=\textbf{C}^+\oplus \textbf{C}^-$. This allows us to define the self-dual manifolds as the ones with $\textbf{C}^-=0$. But lemma \ref{Lemma-Self-dual-man.} guarantees that if $n$ is odd or $n=4$ then the condition $\textbf{C}^-=0$ implies that the whole Weyl tensor vanishes. It was also proved that the integrability condition of a maximally isotropic distribution is given by the invariance of certain subbundles of $\wedge^nM$ under the operator $\textbf{C}_n$.

In theorem \ref{Theor. Optical Scalars} it was shown that if a manifold, of arbitrary dimension and signature, admits a null $p$-form that is closed and co-closed then the optical matrices of the null directions ``tangent'' to this form obey to several constraints. This was then used to arrive at a generalized version of the Goldberg-Sachs theorem stating that if an even-dimensional manifold admits an integrable maximally isotropic distribution then the shear matrix is constrained.

The objects and tools introduced here deserve further investigation and probably the elegant notation introduced in section \ref{Sec- SimpleNot} will help on this enterprise. On theoretical grounds it is important to search for new relations between the operators $\textbf{C}_p$ and integrability properties. This can shed light on the integration of higher-dimensional Einstein's equation, as happened in four dimensions \cite{typeD - Kinnersley}. Also the practical implications of theorem \ref{Theor. Optical Scalars} and of the generalized Goldberg-Sachs theorem should be analyzed more deeply. These theorems should be particularly useful in situations where there are physical fields represented by $p$-forms whose equations of motion implies that they are harmonic (closed and co-closed).

\section*{Acknowledgments}
Thanks to the Conselho Nacional de Desenvolvimento Cient\'{\i}fico e Tecnol\'{o}gico (CNPq) for the financial support. I also want to thank the anonymous referee of the Journal of Mathematical Physics for some valuable suggestions.
\\
The published version is available on \verb"http://dx.doi.org/10.1063/1.4802240"

\appendix
\section{Refined Segre Classification}\label{Appendix-Segre}
In this appendix it will be presented the refined Segre classification, a classification scheme for square matrices over the complex field that was defined in \cite{Spin6D}.

Given a square matrix $M$ over the complex field then by means of a similarity transformation it can always be put in the so called Jordan canonical form. This canonical form is a block-diagonal matrix such that each block is equal to a single number or to a matrix of the following form:
\begin{equation*}
 J=\left[
   \begin{array}{cccc}
    \lambda & 1 & 0 &\ldots\\
                                                                             0 & \lambda & 1&  \\
                                                                             \vdots&  & \ddots &1\\
                                                                             0 & \ldots & 0 &   \lambda \\
   \end{array}
 \right].
\end{equation*}
The Jordan canonical form of a matrix is unique up to the ordering of the Jordan blocks. Each block of the above form has only one eigenvector and its eigenvalue is $\lambda$.

The Segre classification of the matrix $M$ is the list of the dimensions of the Jordan blocks. This list of numbers is put inside a square bracket and the dimensions of the Jordan blocks with the same eigenvalue are to be put together inside a round bracket. This classification can be refined if the numbers corresponding to the blocks with eigenvalue zero are made explicit by putting them on the right of the others and separating by a vertical bar. For example, suppose that the Jordan canonical form of a matrix $M$ is
 \begin{equation}\label{MatrixApp}
 \left[
    \begin{array}{ccccc}
      0& 1 & 0 & 0 & 0 \\
      0 & 0 & 0 & 0 & 0 \\
      0 & 0 & \alpha & 0 & 0 \\
      0 & 0 & 0 & \beta & 1 \\
      0 & 0 & 0 & 0 & \beta \\
    \end{array}
  \right]\,.
 \end{equation}
Then the next table summarizes the possible types that the matrix $M$ can have on the Segre classification and on the refined version of this classification.
\begin{table}[!htbp]
\begin{center}
\begin{tabular}{|c|c|c|c|c|c|}
  \hline
 & \,$0\neq\alpha\neq\beta\neq0$\, &\, $\alpha=0\neq\beta$ \,& \,$\alpha=\beta\neq0$ \,&\, $\beta=0\neq\alpha$ \,& \, $\alpha=\beta=0$ \\
\hline
Segre Classification & $[2,2,1]$ & $[(2,1),2]$ &  $[(2,1),2]$ & $[(2,2),1]$ & $[(2,2,1)]$
\\ \hline
Refined Segre Class. & $[2,1|2]$ & $[2|2,1]$ &  $[(2,1)|2]$ & $[1|2,2]$ & $[|2,2,1]$\\
\hline
\end{tabular}
\caption{Possible algebraic types for the matrix of equation (\ref{MatrixApp}).}
\end{center}
\end{table}

\section{Simple Forms and the Integrability of Distributions}\label{Appendix-SimpleForms}
Let $\textbf{K}_p$ be a $p$-form that is simple, \textit{i.e.}, there exists $1$-forms $\omega^j$ such that $\textbf{K}_p =\omega^1\wedge\omega^2\wedge \ldots \wedge \omega^p $. Using equation (\ref{Hdge Dual}) it is possible to see that the Hodge dual of a simple form is a simple form, so that we can write:
\begin{equation}\label{Dual of H}
   \widetilde{\textbf{K}}_p \;=\; \widehat{\omega}^1\wedge\widehat{\omega}^2\wedge \ldots \wedge \widehat{\omega}^{d-p} \,.
\end{equation}
By means of the metric $g$ it is possible to make a one-to-one association between vectors and 1-forms. For example, the vector field $V$ is associated to the 1-form $\omega$ when $\omega(X)=g(V,X)$ for all vector fields $X$. Using this correspondence let us denote by $V_j$ the vectors associated to the 1-forms $\omega^j$ and by $\widehat{V}_r$ the vectors associated to $\widehat{\omega}^r$. Now let us define the following vector field distributions:
\begin{equation*}\label{N,*N spaces}
  \mathcal{N}=\{V_1,V_2,\ldots,V_p\} \quad\quad; \quad\quad \widetilde{\mathcal{N}}=\{\widehat{V}_1,\widehat{V}_2,\ldots,\widehat{V}_{d-p}\}\,.
\end{equation*}

Using equations (\ref{Hdge Dual}), (\ref{Dual of H}) and the definition $V_j^{\;\mu}=\omega^{j\,\mu}$ it follows that:
$$[V_j\lrcorner(\widehat{\omega}^1\wedge \ldots \wedge \widehat{\omega}^{d-p})]_{\mu_2\ldots\mu_{d-p} } \,=\,\frac{n!}{p!}\, V_j^{\;\mu_1}\varepsilon^{\nu_1\ldots\nu_{p}}_{\phantom{\nu_1\ldots\nu_{p}}\mu_1\ldots\mu_{d-p}} V_{1\,\nu_1}\ldots V_{p\,\nu_p} \,=\, 0\,, $$
where $(V\lrcorner \mathbf{F})$ is the interior product of the vector field $V$ into the differential form $\mathbf{F}$.  This implies that $\widehat{\omega}^r(V_j)=0$, which is equivalent to $g(\widehat{V}_r,V_j)=0$. So we arrived at the following important result:
\begin{equation}\label{N Orthog *N}
  \widetilde{\mathcal{N}} \;=\; \mathcal{N}^\bot\,.
\end{equation}
Where $\mathcal{N}^\bot$ is the orthogonal complement of $\mathcal{N}$.

Given a simple $p$-form $\textbf{K}_p$,  the vector distribution annihilated by it is defined by  $\mathcal{A}_{\mathbf{K}_p} = \{U\,\in\,TM\,|\,U\lrcorner\textbf{K}_p=0\}$.  Since $\widehat{\omega}^r(V_j)=0$ it follows that $\mathcal{N}$ is the vector distribution annihilated by the form $\widetilde{\textbf{K}}_p$, while $\widetilde{\mathcal{N}}$ is the distribution annihilated by $\textbf{K}_p$. It was proved in section \ref{Sec- Mariot-Rbinson} that the distribution $\mathcal{A}_{\mathbf{K}_p}$ is integrable if, and only if, there exists some function $h\neq0$ such that $\textbf{d}(h\textbf{K}_p)=0$, therefore we can state:
\begin{equation*}\label{N integrab}
  \left\{
    \begin{array}{ll}
      \mathcal{N}\quad \textrm{is integrable}\quad \Leftrightarrow \quad \exists \; f\neq0 \;| \;\textbf{d}(f\widetilde{\textbf{K}}_p)=0   \\
      \widetilde{\mathcal{N}}\quad \textrm{is integrable}\quad \Leftrightarrow \quad \exists \; g\neq0 \;\,| \;\textbf{d}(g\textbf{K}_p)=0\,.
    \end{array}
  \right.
  \end{equation*}

In the special case in which $d=2n$, $p=n$ and $\textbf{K}_n$ is a null $n$-form, \textit{i.e}, the distribution $\mathcal{N}$ is maximally isotropic, it follows that $\mathcal{N}^\bot=\mathcal{N}$. So by equation (\ref{N Orthog *N}) we see that $\widetilde{\textbf{K}}_n\propto\textbf{K}_n$, which implies that $\textbf{K}_n$ must be self-dual or anti-self-dual. In this case we have that $\mathcal{N}$ is integrable if, and only if, there exists some function $h\neq0$ such that $\textbf{K}'_n=h\textbf{K}_n$ obeys to the equations $\textbf{d}(\textbf{K}'_n)=0$ and $\textbf{d}(\widetilde{\textbf{K}'}_n)=0$.

\end{document}